\documentclass[]{aa}
\usepackage[]{psfig}
\usepackage{txfonts}
\begin{document}

\title{The He-shell flash in action: T Ursae Minoris revisited}

\author{K. Szatm\'ary \and L. L. Kiss \and Zs. Bebesi} 

\institute{Department of Experimental Physics and Astronomical Observatory,
University of Szeged,
Szeged, D\'om t\'er 9., H-6720 Hungary}

\titlerunning{T~UMi revisited}
\authorrunning{K. Szatm\'ary et al.}
\offprints{k.szatmary@physx.u-szeged.hu}
\date{}

\abstract{We present an updated and improved description of the light curve 
behaviour of T Ursae Minoris, which is a Mira star with the
strongest period change (the present rate is an amazing $-3.8\pm$0.4 
days/year corresponding to a relative decrease of about 1\% per cycle). 
Ninety years of visual data were collected from all
available databases and the resulting, almost uninterrupted light curve 
was analysed
with the O$-$C diagram, Fourier analysis and various time-frequency methods.
The Choi-Williams and Zhao-Atlas-Marks distributions gave the clearest 
image of frequency and light curve shape variations. A decrease of the
intensity average of the light curve was also found, which is in accordance 
with a period-luminosity relation for Mira stars.
We predict the star will finish  
its period decrease in the meaningfully near future (c.c. 5 to 30 years) and 
strongly suggest to closely follow the star's variations (photometric, as well
as spectroscopic) during this period.
}
 
\maketitle

\section{Introduction}

During the last several years, there has been an increasing number of 
Mira stars discovered to show long-term continuous period changes (see
some recent examples in Sterken et al. 1999, Hawkins et al. 2001 
and a comprehensive reanalysis of R~Hydrae, an
archetype of such Mira stars, by Zijlstra et al. 2002). The widely 
adopted view of their period change is based on the 
He-shell flash model, outlined by Wood \& Zarro (1981). According to this
model, energy producing instabilities appear when a helium-burning shell, 
developed in the early Asymptotic Giant Branch (AGB) phase, starts to 
exhaust its helium content. Then the shell switches to hydrogen burning, 
punctuated by regular helium flashes, called also as thermal pulses
(Vassiliadis \& Wood 1993). During these flashes the stellar luminosity 
changes quite rapidly and the period of pulsation follows the luminosity 
variations.

Rapid period decrease in T Ursae Minoris (=  HD~118556, V$_{\rm max}\approx
9\fm0$, V$_{\rm min}\approx14\fm0$) was discovered by 
G\'al \& Szatm\'ary (1995), who analysed $\sim$45 years-long visual data
distributed in two distinct parts between 1932 and 1993. Mattei \& Foster (1995)
analysed almost 90 years of AAVSO data collected between 1905 and 1994 and 
concluded that the period decreasing rate of T~UMi (2.75 days/year as
determined by them) is twice as fast as in two other similar Mira stars (R~Aql
and R~Hya). Most recently, \v{S}melcer (2002) presented almost three years of
CCD photometry of T~UMi, resulting in four accurate times of maximum. 

The main aim of our paper is to update our knowledge on T~UMi. 
The eight years passed since the last two detailed analyses witnessed a 
considerable development in time-frequency methods, which allow more
sophisticated description of light curve behaviour. On the other hand, these 
eight years yielded more than 10 new cycles of the light curve prolonging
quite significantly the time-base of the period decreasing phase.
The paper is organised as follows. Observations are described 
in Sect.\ 2, a new and more sophisticated light curve analysis is presented
in Sect.\ 3. Results are discussed in Sect.\ 4.

\section{Observations}

Four sources of visual data were used in our study.
The bulk of the data was taken from the publicly available databases 
of the Association Fran\c caise des Observateurs d'Etoiles Variables
(AFOEV\footnote{\tt ftp://cdsarc.u-strasbg.fr/pub/afoev}) and
the Variable Star Observers's League in Japan 
(VSOLJ\footnote{\tt http://www.kusastro.kyoto-u.ac.jp/vsnet/gcvs}).
(Besides the visual data, the AFOEV subset contains a few CCD-V measurements,
too.) Since these data end in early 2002, the latest 
part of the light curve is covered via the VSNET 
computer service. The merged dataset showed a quite large gap between 
JD 2431000 and 2437000. Therefore, we have extracted data collected 
by the American Association of Variable Star Observers (AAVSO) with help
of the Dexter Java applet available at the Astrophysical Data System (these
data were published by Mattei \& Foster (1995) as light curves, which can be 
converted into ASCII data with that Java applet). 
Basic data of individual sets are summarized in Table\ 1.

\begin{table}
\begin{center}
\caption{A summary of the analysed datasets (MJD=JD$-$2400000).}
\begin{tabular}{|lllr|}
\hline
Source & MJD(start) & MJD(end) & No. of points\\
\hline
AFOEV  & 22703 & 52457 & 4073\\
AFOEV(CCD) & 51071 & 52321 & 271\\
VSOLJ  & 36691 & 52273 & 891\\
VSNET  & 49925 & 52551 & 501\\
AAVSO$^a$ & 20043 & 49530 & 3213\\
\hline
\end{tabular}
\end{center}
$^a$ Mattei \& Foster (1995)
\end{table}

We found the different data to agree very well (similarly to the case of 
R~Cygni in Kiss \& Szatm\'ary 2002) and that is why we simply merged
the independent observations to form the final dataset. It is almost
uninterrupted between 1913 and 2002 and consists of 8949 individual magnitude
estimates (negative (``fainter than...'') observations were excluded). 
We have calculated 10-day means and this binned light curve was
submitted to our analysis. It is shown in Fig.\ 1, where the close agreement 
of simultaneous CCD-V and visual observations supports the usability of 
the latter data (some tests and comparisons of photoelectric and 
visual observations can be found in Kiss et al. 1999 and Lebzelter \& Kiss
2001). 

\begin{figure}
\begin{center}
\leavevmode
\psfig{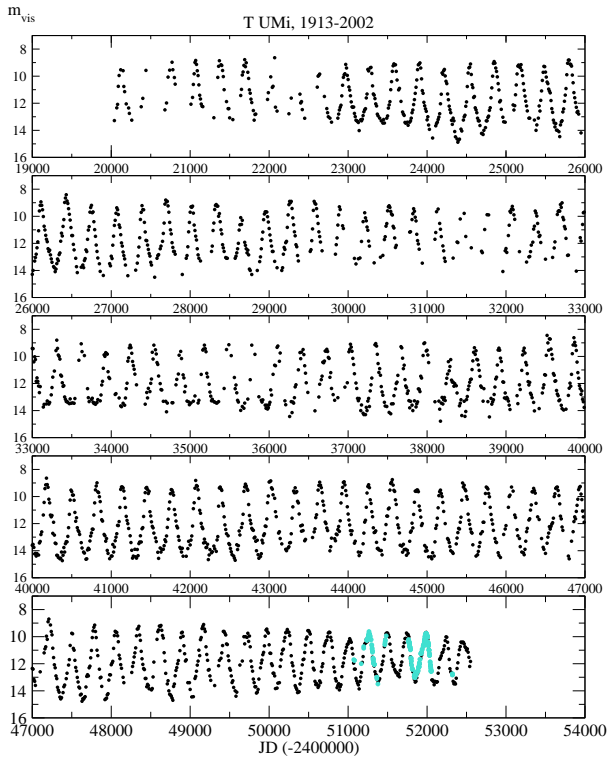}
\caption{The whole dataset of T~UMi (10-day means). Light gray points in
the bottom panel refer to 
CCD-V measurements collected by the AFOEV.}
\end{center}
\label{fig1}
\end{figure}

\section{Light curve analysis}

\subsection{The $O-C$ diagram}

To construct the classical O$-$C diagram, we determined all times of maximum
from the binned light curve. This was done partly by fitting low-order (3--5)
polynomials to selected parts of the light curve, partly by simple 
`eye-ball' estimates of the epochs of maximum from computer generated plots.
The latter was chosen when the scatter of the curve and/or loose sampling
did not permit reliable fitting. We could determine 106 observed epochs
(with estimated errors of $\pm$5 and $\pm$10 days), i.e. 
only very few cycles were lost between 1913 and 2002. Eight additional 
times of maximum was provided by J. Percy (1994, personal communication),
which were observed between 1907 and 1913. Therefore, the full set consists of
114 observed maxima with a time-base of 95 years. To allow an easy 
comparison with Fig.\ 4 in G\'al \& Szatm\'ary (1995), we plot the resulting
O$-$C diagram with the same ephemeris ($JD_{\rm max}=2416971+313.42\cdot E$)
in Fig.\ 2. 

Obviously the star continued the period decrease at an amazing rate. A close-up
to the last part of the O$-$C diagram revealed its very parabolic nature
indicating constant rate of period change (similar conclusion was drawn by
\v{S}melcer 2002). By fitting a parabola to the last 7500 days, 
we determined the second-order coefficient as 
$(-1.82\pm0.18)\cdot10^{-5}$ days/days. Using the expression for this
coefficient in terms of period and period changing rate
($\frac{1}{2}\frac{1}{P}\frac{dP}{dt}$, Breger \& Pamyatnykh 1998),
we obtained a relative rate of $(1/P)dP/dt=(-3.6\pm0.36)\cdot10^{-5}$ days/days.
Here $P$ stands for the period in the ephemeris used to construct the O$-$C
diagram, hence the period derivative $dP/dt$ equals to $-4.2\pm0.4$ days/year.

\begin{figure}
\begin{center}
\leavevmode
\psfig{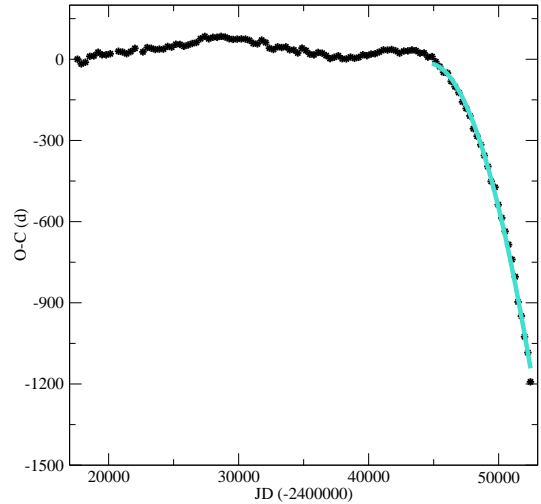}
\caption{The O$-$C diagram of T~UMi. The solid line is the parabolic fit 
of the last 7500 days.}
\end{center}
\label{fig2}
\end{figure}

\begin{figure}
\begin{center}
\leavevmode
\psfig{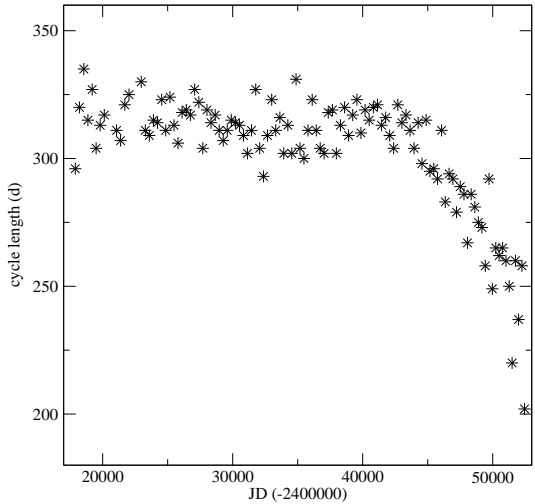}
\caption{The cycle length as a function of time.}
\end{center}
\label{fig3}
\end{figure}

The rate of period decrease was also determined from the individual cycle
lengths. They are plotted as a function of time in Fig.\ 3. There is a 
well-expressed break at JD 2444000 and since then the period has decreased 
fairly linearly, although some curvature cannot be excluded either. 
By fitting a linear regression to the last 7500 days, we arrived to 
a similar result: $dP/dt=3.4\pm0.5$ days/year. This is about the same 
than that of \v{S}melcer (2002) who gave 3.1 days/year. 
In summary, we adopt the simple mean of the two values which is 
$dP/dt=3.8\pm0.4$ days/year. Presently it corresponds to about 
1 percent relative decrease per cycle! To our knowledge, there is 
only one star with similarly fast gradual period change among all types
of pulsating stars (sudden period changes due to mode switching phenomenon 
are not considered here): BH~Crucis, which is another Mira star but with 
period increase (Zijlstra \& Bedding 2002). Its period has increased since 1975
from 420 to 530 days, i.e. with an approximate rate of about 
4 days/year. Whatever the reason is, these stars are obviously very
peculiar members of their class (see Percy \& Au 1999 for a search
for evolutionary period changes in 391 Mira stars).

\subsection{Fourier analysis and time-frequency distributions}

Fourier and time-frequency analyses make use of the full light curve,
not only special points, thus their application usually reveals a lot more
information on the light variation. 

\begin{figure}
\begin{center}
\leavevmode
\psfig{figure=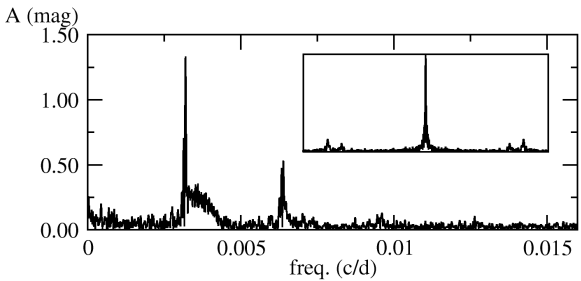,width=8cm}
\caption{The frequency spectrum of T~UMi with the window function in the small
insert. The frequency scale is the same for both graphs.}
\end{center}
\label{fig4}
\end{figure}

First, we calculated the frequency spectrum of T~UMi with Period98 (Sperl 1998).
It is plotted in Fig.\ 4, where the window function is also shown in the small
insert. The shape of the spectrum around the dominant peak is very
characteristic for a signal with continuously changing frequency. The second and
third harmonics (due to the asymetric light curve shape) are easily visible 
and there is a slight indication for the fourth harmonic, too. Thanks to the
almost uninterrupted light curve, the window function is free of strong alias
structure, only the weak yearly alias and a stronger $1/P$ alias are present 
(the latter one is caused by a few unobserved minima in the early decades
of data). 

However, as has been recently demonstrated by numerous studies 
(e.g. Szatm\'ary et al. 1996, Foster 1996, Koll\'ath \& Buchler 1996, 
Bedding et al. 1998, Kiss et al. 2000, Zijlstra et al. 2002), time-frequency
analysis utilizing wavelets and other distributions (Cohen 1995) reveals many
delicate details undetectable with simple methods.

\begin{figure*}
\begin{center}
\leavevmode
\caption{Three time-frequency distributions. Top panel: wavelet map;
middle panel: Choi-Williams distribution (CWD); bottom panel: 
Zhao-Atlas-Marks distribution (ZAMD).}
\end{center}
\label{fig5}
\end{figure*}

We calculated several distributions with many different parameter sets 
with the software package TIFRAN (TIme FRequency ANalysis) developed 
by Z. Koll\'ath and Z. Csubry at Konkoly Observatory, Budapest (Koll\'ath \&
Csubry 2002). Besides the 
wavelet map, we present the Choi-Williams (Choi \& Williams 1989)
and Zhao-Atlas-Marks (Zhao et al. 1990) time-frequency distributions
for the whole dataset of T~UMi in three panels of Fig.\ 5. For an easy
comparison we also plotted the light curve at the top of the figure and
three copies of the Fourier spectrum on the right hand side of the
distributions.
In order to enhance the visibility of harmonic components, we 
divided the time-frequency plane in three sections defined by their 
frequency limits. We chose 0--0.0048 d$^{-1}$, 0.0048--0.0096 d$^{-1}$
and 0.0096--0.015 d$^{-1}$. This selection provided that the fundamental
frequency ($f_0$), and its second ($2f_0$) and third ($3f_0$) harmonics 
fall into different sections. Then we multiplied the amplitude values 
in these sections by one, two and three, with the larger frequency 
the larger multiplicator. Detailed experiments were performed with 
different kernel-parameters $\alpha$ (Buchler and Koll\'ath 2001),
which defines the tradeoff between time resolution and frequency 
resolution. In the presented cases for the wavelet transform, CWD 
and ZAMD we used $\alpha$ values of 2.0, 2.0 and 0.2, respectively.

We could draw a few interesting conclusions based on Fig.\ 5. Both the CWD and
ZAMD give much clearer images of the time-dependent frequency content 
than the wavelet map. Even the fourth harmonic is visible, although with 
some ambiguity. The light curve shape (defined by the relative strength of
the harmonics) changed a lot even in the first part of the data, in which the
period remained constant. When the period decrease started (around JD 2444000
or 1979), 
the harmonics followed the fundamental frequency for $\sim$5000 days but 
around JD 2449000 they suddenly disappeared -- the star 
became ``tuned out''. Consequently, the light curve shape turned to be
more sinusoidal. Between 1913 and 1979 we did not find any significant 
frequency change, only the amplitude of the fundamental and first harmonic
seems to show some alternating changes. In order to decide whether they are
real
changes or only results of the data distribution, we performed a similar 
test calculation as in Szatm\'ary et al. (1996). It consisted of 
creating a wavelet map of the original dataset (`star') and another map
from a synthesized light curve (`fit') calculated with the two dominant
frequency components in the Fourier spectrum. The fit-map shows the amplitude
variations caused by the gaps and data distribution. A comparison of the
star-map with the fit-map (in sense of star minus fit) reveals real amplitude
variations.

We show the results in Fig.\ 6. Here A$_{\rm i}$ (i=0,1) stands for 
the star$-$fit difference for the peak amplitudes in the corresponding
frequency ranges at any given time. The decrease of the first harmonic 
after the start of the period decline is quite obvious. There are several
occasions when A$_0$ and A$_1$ showed alternation but that was not a 
strict rule. If the star was pulsating in fundamental mode, then
the reason for this complex behaviour might be some resonance
between the fundamental and first overtone modes, for which a ratio near
2.0 is expected in red giants models (e.g. Ostlie \& Cox 1986). 

\begin{figure}
\begin{center}
\leavevmode
\psfig{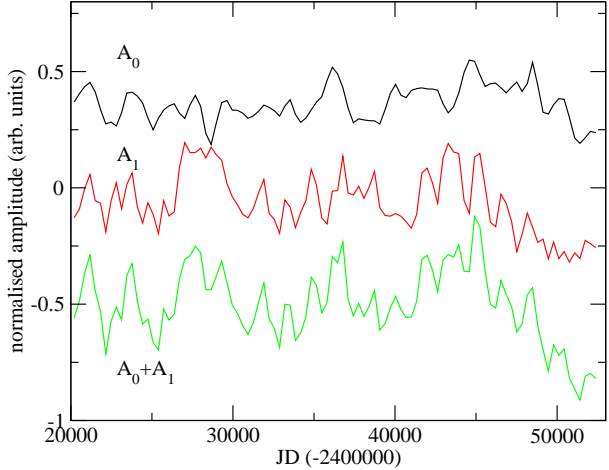}
\caption{Amplitude variations of the fundamental (A$_0$) and first harmonic 
frequencies (A$_1$).}
\end{center}
\label{fig6}
\end{figure}

\section{Discussion}

We have investigated several aspects and implications of the presented 
light curve behaviour. An intriguing question is whether the overall luminosity
drop expected from the assumption of a period-luminosity (P-L) relation  
can be detected. This question was addressed by Zijlstra et al. (2002) for 
R~Hydrae, whose period decreased from 495 days to 380 days between 1700 and
1950. These authors examined different assumptions, including luminosity 
decrease from a P-L relation (Feast 1996), evolution at constant $L$ 
and even evolution with slightly increasing $L$. From the visual light curve 
it was concluded
that no luminosity change can be proven as the average visual magnitude 
has not changed since 1910. However, they noted the decreased visual amplitude
explained by the non-linearity of pulsation (cf. Kiss et al. 2000 discussing
the period-amplitude relation for Y~Persei). A similar amplitude decrease was
described for another He-shell flash Mira, R Centauri (Hawkins et al. 2001).

However, we feel it necessary to point out that for a large amplitude 
Mira star one has to be careful when concluding the luminosity constancy from
the constant average magnitude. This is because of the logarithmic nature of
the magnitude scale. As an illustration, let us consider two Mira stars with 
$m_{\rm 1,max}=7$ mag, $m_{\rm 1,min}=13$ mag, $m_{\rm 2,max}=8$ mag 
and $m_{\rm 2,min}=12$
mag. The average magnitude is 10 mag in both cases. However, the physically
relevant parameter is the intensity being $i\sim10^{-0.4 m}$.
After calculating average intensities, one can convert them to meaningful 
average magnitudes, in our cases to 7\fm75 and 8\fm73. The difference is 
almost one magnitude. This is a fairly trivial consideration but it has
to be kept in mind when averaging large amplitude Mira light curves.

As expected, both R~Hya and R~Cen showed such amplitude decrease that resulted
in fainter maxima and brighter minima (Figs.\ 1 in Hawkins et al. 2001 and
Zijlstra et al. 2002). We found a similar behaviour for T~UMi, too, which 
has already been predicted by Whitelock (1999).
To take into account the necessity of intensity averaging, we have converted 
the binned light curve to intensities and calculated mean intensities per
cycles (Fig.\ 7). The scatter in Fig.\ 7 is, of course, quite large, due 
to typical cycle-to-cycle changes of the maximum brightness. However, 
there is an indication at the end of data for a slight decrease of the 
intensity. A simple linear fit of the last 7500 days (JD 2445000--2452500) 
was used to infer $\Delta \langle m \rangle=-0\fm47\pm0\fm4$. 
This value was compared with calculated absolute magnitude changes (see next
paragraph).
The general appearance of the intensity average curve as a function of time 
reminded us the shape of the O$-$C diagram and that is why we added the
O$-$C points (before JD 2445000) in Fig.\ 7. The parallel trends
suggest that there is a positive correlation between the full intensity and 
corresponding cycle length. 

As a rough approximation, we assumed that averaging along the full cycle
smooths out the effects of varying bolometric correction and the mean magnitude
difference found from the observations corresponds directly to 
$\Delta M_{\rm bol}$. Adopting a period change from 315 days to 215 days, 
the P-L relation derived from LMC Mira stars (Feast 1996)

$$M_{\rm bol}=-3.00 \log P + \alpha$$

\noindent gives $\Delta M_{\rm bol}=-0\fm49$. The close agreement
is probably pure coincidence, though remarkable.

Another comparison was made with theoretically expected luminosity change as
discussed in Wood \& Zarro (1981) when deducing their Eq. (10). From that
relation it follows that the luminosity change depends only on the period
change and two ambiguosly determined constants ($b$ and $\beta$ in their
notation). For $\beta=16.67$ and $b=1.5$ or $2$, we calculated 
$\Delta M_{\rm bol}=-0\fm45$ and $-0\fm34$. The agreement is again 
demonstrative. Therefore, we conclude that the tenuous intensity decrease 
is consistent (at least partly) with the assumption of P-L relation being
applicable in this case. Another effect with similar outcome is the amplitude
reduction due to the non-linearity of pulsation. The most likely scenario is
that both processes occur in T~UMi.

Fig.\ 3 in Wood \& Zarro (1981) led G\'al \& Szatm\'ary (1995) to 
conclude that T~UMi is just after the onset of a He-shell flash and it is 
likely to have core mass similar to R~Aql and R~Hya. 
Presently available data shift the core mass to a slightly larger value
(between 0.69 and 0.78 $M_\odot$), because the most recent $dL/dt$ 
suggest a steeper function than that for R~Hya in Fig.\ 3 of Wood \& Zarro 
(1981). Adopting those calculations, we estimated some basic parameters 
of the star. The logarithm of the luminosity for the larger core mass is
about 4.2$\pm$0.1 (i.e. $L/L_\odot=15800^{+4000}_{-3300}$). That means 
$M_{\rm bol}=-5\fm8\pm0\fm3$ corresponding to a spectral type of
M4~II with $M_{\rm V}=-3\fm1\pm0.3$ (Strai\v{z}ys \& Kuriliene 1981).
This results in a distance of 3.6$\pm$0.5 kpc
(interstellar reddening neglected). The definition of luminosity and the
assumption of a typical red giant temperature $T_{\rm eff}=3000\pm200$ K
yields $R/R_\odot=450\pm80$. This large radius suggests first overtone
pulsation for T~UMi (van Belle et al. 2002). On the other hand, period-gravity
relation of radially pulsating stars (Fernie 1995) and its extension toward 
red giant pulsators, mainly semiregular variables 
(Szatm\'ary \& Kiss 2002) suggest for fundamental and first overtone
pulsation $\log~g=-0.22$ and $-0.52$, respectively. The corresponding masses 
are 4.6 $M_\odot$ and 2.3 M$_\odot$, the former being too large for a Mira
star. Tabulated model calculations by Fox \& Wood (1982) for similar
physical parameters and first overtone periods near 310 days give second
overtone periods between 210 and 240 days, close to the presently
observable value. Another possibility is that the continuous period change 
is due to a mode switching phenomenon acting similarly 
than the pop. II models in type-1 instability region calculated by 
Bono et al. (1995). Although those models addressed transient phenomena in
RR~Lyr and BL~Her stars, type-1 instability (mode switching from the
fundamental to the first overtone mode with continuously changing period 
-- see Fig.\ 2 in Bono et al. 1995) is roughly similar to the observed 
behaviour of T~UMi.

Either mode change or He-shell flash is acting in T~UMi, present large period 
changing rate suggests that the decrease will stop in the meaningfully near 
future, between 5 to 30 years from now (i.e. to keep the period in a 
physically reasonable range).
Furthermore,
if the He-shell flash model is true and the star is indeed just after the onset
of a flash, a similarly rapid period increase can be predicted right after 
reaching the period minimum. Therefore, it is of paramount importance to follow
the star's variations with as much instrumentation as possible. The average
cycle length now is between 200 and 220 days and the rapid decrease implies 
that the end of the decline is quickly approaching (Zijlstra et al. 2002 
found for R~Hya that its period showed a c.c. 10\% decrease in two decades
before ending the period changing phase). In this phase we suggest to 
try to measure directly the luminosity change (if we accept its existence)
via accurate spectrophotometry or high-resolution spectral synthesis. Visual
data are crucial for prompt detection of period stabilization or even period 
increase. The latter would be the final argument confirming the concept 
of the He-shell flash. 
However, if the period will turn to a constant value and remains there for a
considerable time then the whole theory should be revised. In that case
T~UMi shall shed new light on a peculiar mode switching phenomenon not 
well understood.

\begin{figure}
\begin{center}
\leavevmode
\psfig{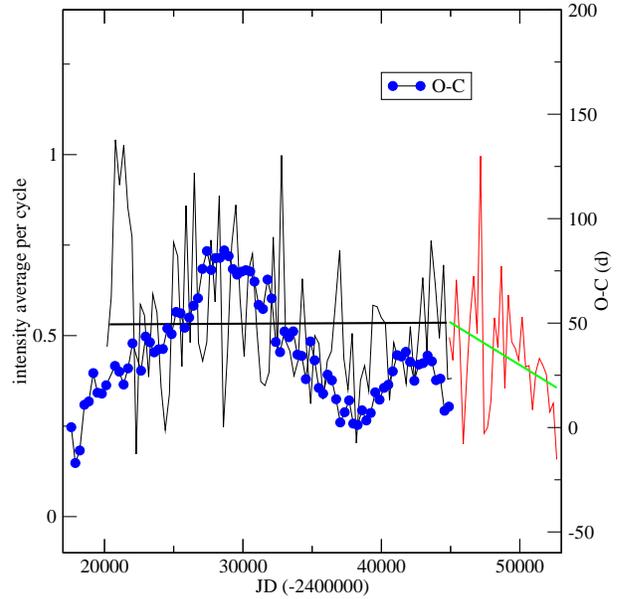}
\caption{The intensity average per cycle against time and
the O$-$C variation before the period decrease.}
\end{center}
\label{fig7}
\end{figure}

\begin{acknowledgements}
This work has been supported by the Hungarian OTKA
Grants \#T032258 and \#T034615, the ``Bolyai J\'anos'' Research 
Scholarship to LLK from the Hungarian Academy of Sciences, 
FKFP Grant 0010/2001 and Szeged Observatory Foundation. 
We sincerely thank variable star observers of AFOEV and VSOLJ 
whose dedicated observations over a century made this
study possible. The computer service of the VSNET group is
also acknowledged. We are grateful to Dr. Z. Koll\'ath for providing
the TIFRAN software package.
The NASA ADS Abstract Service was used to access data and
references. This research has made use of the SIMBAD database, operated at
CDS-Strasbourg, France.
\end{acknowledgements}

\end{document}